\documentclass[a4paper,english,preprint, aps]{revtex4-2}
\usepackage[T1]{fontenc}
\usepackage[latin9]{inputenc}
\usepackage{float}
\usepackage{amsmath}
\usepackage{amssymb}
\usepackage{graphicx}

\makeatletter

\renewcommand*{\p@subsection}{}

\renewcommand*{\p@subsubsection}{}

\makeatother

\usepackage{babel}
\begin{document}
\preprint{Preprint to be Submitted}
\title{CONFORMABLE DERIVATIVE APPROACH TO GRANULAR GASES}
\author{José Weberszpil}
\email{josewebe@gmail.com; josewebe@ufrrj.br}

\affiliation{Group of Theoretical Physics and Mathematical Physics, Department
of Physics, ~\\
Federal Rural University of Rio de Janeiro, ~\\
Cx. Postal 23851, BR 465 Km 7, 23890-000 Seropédica - RJ, Brazil.~\\
}
\author{Cresus F. de L. Godinho}
\email{crgodinho@gmail.com; crgodinho@ufrrj.br}

\affiliation{Group of Theoretical Physics and Mathematical Physics, Department
of Physics, ~\\
Federal Rural University of Rio de Janeiro, ~\\
Cx. Postal 23851, BR 465 Km 7, 23890-000 Seropédica - RJ, Brazil.~\\
}
\author{Ion Vasile Vancea}
\email{ionvancea@ufrrj.br; ivvan2008@gmail.com}

\affiliation{Group of Theoretical Physics and Mathematical Physics, Department
of Physics, ~\\
Federal Rural University of Rio de Janeiro, ~\\
Cx. Postal 23851, BR 465 Km 7, 23890-000 Seropédica - RJ, Brazil.~\\
}
\date{\today}
\begin{abstract}
Proper modeling of complex systems requires innovative mathematical
tools. In this sense, we sought to use deformed derivatives for studying
the dynamics of systems, particularly those, such as granular gases,
in which the description of the dynamics can be also done by using
the stretched exponential probability densities. In this contribution
we draw up three results of this application of mathematical tools.
The first result shows that when we use constraints with finite momentum
and the principle of maximum entropy, the Kohlrausch--Williams--Watts
function, known as stretched exponential, emerges naturally and in
a simpler way, when compared to results in the literature. Next, we
obtain generalized expressions for the Langevin equation, as well
as its solutions for three different deformed derivatives, including
those connected with nonaddictive statistical mechanics. The Haff's-like
law for granular gases are obtained. Next, we calculate the partition
function $Z$ for granular gases systems by building up the probability
density in terms of the stretched exponential function. From this
partition function, we determine the internal energy of the system
as well as the specific heat, both dependent on temperature. The consistency
with classical approach of kinetic theory for ideal gases was verified.
Comparatives with experimental data were taken into account 
\end{abstract}
\maketitle

\section{Introduction}

As an example to describe complex systems characterized by dynamics
with power laws and by non Gaussian Maxwell-Boltzmann probability
distributions, the stretched exponential probability densities ($SEPD$)
or Kohlrausch--Williams--Watts ($KWW$) function can be cast as
one of the most ubiquitous form to analyze a wide range of physical
phenomena, \citep{luevano2013statistical}. This is based on experimental
evidences describing a relaxation of various natures \citep{vlad1996universality,lukichev2019physical,laherrere1998stretched},
including biological one. The nonlinear dissipation of energy signalize
the main point to the use of $KWW$-type relaxations.

In Ref. \citep{vlad1996universality}, the authors conjectured that
some universal mechanism generate $KWW$ dynamics which is independent
of the details of an individual process may play an important hole.
Also, it is possible to weave a connection between the $KWW$ law
and the stable probability densities of the Lèvy type, justified as
a result of occurrence of a large number of independent random events
described by individual probability densities with infinite moments.
The Ref. \citep{gorska2017stretched} can also be consulted for connections
with Lèvy-type relaxation. For several interesting applications and
comments, the reader are suggested to consult the Ref. \citep{maruoka2016shannon}
and references inside, including those about scale and size effect
of the system.

The $KWW$ functions are very present for the description of the dynamics
of granular gases and granular particles. The inelastic collisions
of driven dissipative system and non-equilibrium phenomena are related
to some forms of generalized statistical mechanics. Granular gases
are dilute systems of grains in motion with small volume fraction
of grains and the statistical mechanics involved of can be investigated
analyzing granular gases experiments. There are experimental evidences
that for granular particles, the normalized velocity distributions
deviate from the Gaussian Maxwell-Boltzmann distribution found in
equilibrium \citep{shokef2006thermodynamic} and exhibit stretched
exponential velocity distributions \citep{ben2003inelastic} and can
give rise to collective phenomena like spontaneous formation of cluster
in non-equilibrium dynamical states. It is worth to mention that some
behavior of granular media can be connected to its ability to form
a hybrid state between a fluid and a solid, considered as a multi-particle
system \citep{luding2002structures}.

It is interesting to indicate a recent result, in Ref. \citep{yu2020velocity},
where the authors, using a micro-gravity experiment that allows the
spatially homogeneous excitation of spheres via magnetic fields, obtained
a nonlinear fit of the experimental data using stretched exponential
that showed a parameter $\alpha$ near $1$ (here we call this case
a low level fractionality).

A model for vibranting granular gases was also considered in Ref.
\citep{demaerel2020producing} giving rise to stretched exponential
tails.

To tackle the study of alternative description the dynamics of granular
gases, here we use theoretical approaches that are non-standard. We
note that some classical systems are not well described by the Maxwell-Boltzmann-
Gibbs statistic, so, this can leads to a generalized statistical mechanics.
Complementary to the context of non-additive statistics, there are
some local operators, as the Hausdorff derivative \citep{chen2006time},
intimately connected with the conformable derivative ($CD$) \citep{KHALIL201465,abdeljawad2015conformable}
and some more general extensions, like the general conformable derivatives
\citep{zhao2017general} and structural derivatives \citep{xu2017spatial}.
These local operators indicate to be capable of describing complex
systems dynamic behaviors such as anomalous diffusion, creep and relaxation
in fractal media and also to overcome high computing costs of the
non-local fractional derivative \citep{chen2017non}. Another local
deformed derivative operator, the q-derivative emerges from generalized
statistical mechanics \citep{borges2004possible}, connected with
Hausdorff derivative \citep{weberszpil2015connection,sotolongo2021explicit}.
For the sake of clarity, we can encompass all these local derivatives
in a general nomenclature that we will call deformed derivatives ($DD$).

In order to better clarify the connection and justification of the
use of this approach, we can say that the Hausdorff derivative is
defined within an euclidean geometry but with a spatial fractal metric
\citep{balankin2012hydrodynamics,balankin2012map,balankin2013electromagnetic,balankin2016towards}
or with a fractal time metric \citep{yang2018local}. We can also
say this for other $DD$ operators. Through the mapping to a continuous
fractal \citep{balankin2012map,balankin2012hydrodynamics}, these
derivatives can be related to the flux of a fluid in a fractal medium.
The cost for this mapping is that it can occur certain deformation
in the derivative describing the dynamical evolution of the system
and that can lead to a generalized statistical physics. Note that
even the non-local derivatives, like the Riemann Liouville fractional
derivative can be conceived as a global structural derivative \citep{su2018non}.
The expression of the Hausdorff derivative is identical to that given
in eq. (\ref{eq:conformable-deriv-differentiable}). For more details
relating $DD$ and non-Euclidean metric, the readers can consult Ref.
\citep{chen2017non}. Also, for the definition of stretched exponential
stability the reader can consult the Ref. \citep{chen2018stretched}.
Concerning stretched exponential from generalized statistical mechanics,
e.g., the Tsallis statistics, the reader can consult Ref. \citep{beck2006stretched},
where applications for granular gases are presented, indicating the
importance of stretched exponential hole. In that refered article,
the authors claim that a dynamical reason for fat tails can be a so-called
super statistics, where one has a superposition of local Gaussian,
whose variance fluctuates on a rather large spatio-temporal scale.

In Ref. \citep{Cresus2024Remarks} we explore a generalized thermodynamics,
with a quantum version of $C_{V}.$

With all of this in mind, here we can consider that the dynamical
evolution of the granular system follows possible anomalous dynamics,
characterized by different dynamical equations and with the presence
of dissipation intrinsically. By this justification, we generalize
the Langevin Equations ($LE)$, to describe granular gases dynamics
as dissipative systems and, for such intend we consider different
forms of $DD$ as derivatives which are included in the kinetic equations.
As a consequence of this description, the geometry of phase-space,
implicit in the choice of $DD$ by the mapping to fractal continuous,
has deep influence in the form of the solutions for the corresponding
deformed $LE$.

Here we claim, as in Ref. \citep{Jap-Weber-Sotolongo}, that new conceptions
and approaches, such as $DD$, may allow us to understand new systems.
In particular, the use of deformed derivatives (local), similarly
to the (nonlocal) fractional calculus ($FC$), allows us to describe
and emulate complex dynamics involving environmental variation, without
the addition of explicit terms relating to this complexity in the
dynamical equations describing the system, i.e, without explicit many-body,
dissipation or geometrical terms \citep{sotolongo2021explicit}.

Our article is outlined as follows. Section 2 addresses some brief
comment and introduction about granular gases. In Section 3, we focus
on the mathematical aspects of deformed derivative operators. In Section
4, some breaf comments about maximum entropy principle, using stretched
exponential distribution and the influence of constrains is taken
into account. In Section 5, one develop models of Langevin equation
with deformed derivative operators. In Section 6, a temperature dependent
specific heat for the stretched exponential distribution is developed.
Finally, in Section 7, we present our general conclusions and possible
paths for further investigations. 

\section{Granular Gases}

Granular gas are complex systems, in non-equilibrium state and composed
of macroscopic particles \citep{wang2010experiments} where energy
dissipation is governed by binary inelastic collisions between particles
or granules. The system is dilute and particles interact with each
other through instantaneous inelastic processes, so that the interactions
are a continuous sink of kinetic energy. Granular systems exhibit
solid, liquid or gas-like behavior, depending on the external condition
owing to the difference in the dominant physical process of energy
dissipation with many interesting behaviors. A non-equilibrium steady
state can be sustained by a continuous energy injection (steady state)
that if it is stopped, the system is brought into a freely evolving
state \citep{tatsumi2009experimental}. 

In Ref. \citep{tatsumi2009experimental}, under microgravity conditions,
the frictional force was reduced and the authors claims to confirm
that the velocity distribution function has the form $e^{(-\alpha|v|^{\beta})}$.
The value of exponent $\beta$ depends on the conditions given in
the reference. They also relates that in certain conditions the system\textquoteright s
energy decays algebraically $\left(Tg=\frac{T0}{(1+t/\tau)^{\gamma}}\right)$,
agreeing with Haff\textquoteright s law \citep{haff1983grain}.

The statistical physics of granular gases is unusual and give rise
to pronounced deviation from classical Boltzmann statistics, exhibiting
interesting collective phenomena \citep{ben2005power,herrmann2001dynamics}.
To study the dynamics of such systems, we claim that the $DD$ approach
can be an alternative simple one, for naive models.

\section{Mathematical Aspects}

\subsection{Deformed Derivatives}

Here we present some mathematical aspects as in Ref. \citep{sotolongo2021explicit}
for the the sake of clarify.

The conformable derivative ($CD$) is defined as \citep{KHALIL201465} 

\begin{equation}
D_{x}^{\alpha}f(x)=\lim_{\epsilon\rightarrow0}\frac{f(x+\epsilon x^{1-\alpha})-f(x)}{\epsilon},\label{eq:Deformed Derivative-Deff}
\end{equation}
with $0<\alpha\leq1.$

Note that the deformation is given in the independent variable. This
is a local operator.

With a simple change of variable operation, $h=\epsilon x^{1-\alpha}$,
$CD$ can be written for differentiable functions as,

\begin{equation}
D_{x}^{\alpha}f=x^{1-\alpha}\dfrac{df}{dx}.\label{eq:conformable-deriv-differentiable}
\end{equation}

Connected with $CD$, the conformable integral is a Riemann improper
integral \citep{leopoldino2019discussing} 
\begin{equation}
I_{x}^{\alpha}f(x)=\int_{0}^{x}x'^{\alpha-1}f(x')dx',
\end{equation}
 in such a way that a fundamental theorem can be easily proved \citep{KHALIL201465,abdeljawad2015conformable}:
\begin{equation}
D_{x}^{\alpha}I_{x}^{\alpha}f(x)=f(x).\label{eq:conformable Fundamental Theorem}
\end{equation}

Also a local operator is the dual conformable derivative ($DCD$),
which was introduced in Ref. \citep{rosa2018dual} as the dual of
$CD$.
\begin{equation}
\tilde{D}_{x}^{\alpha}f=f^{\alpha-1}\dfrac{df}{dx},
\end{equation}

Here, the deformation is related to the dependent variable.

A simple basic eigenvalue equation emerges and is given by the eq.(34)
in ref. \citep{rosa2018dual}, as

\begin{equation}
[f(x)]^{\alpha-1}\dfrac{df(x)}{dx}=f(x).\label{eq:dual-eigen equation}
\end{equation}
Solving the equation (\ref{eq:dual-eigen equation}), along with the
condition $F(0)=1,$ leads to the following important function:

\begin{equation}
f(x)=\left[1+(\alpha-1)x\right]^{1/(\alpha-1)}.
\end{equation}

Redefining the relevant parameter $\alpha$ in terms of the entropic
parameter $q,$ as $\alpha=2-q$, the solution to eq. (\ref{eq:dual-eigen equation})
becomes

\begin{equation}
f(x)=\left[1+(1-q)x\right]^{1/(1-q)}=e_{q}(x),
\end{equation}
that is exactly the $q-exponential$, $e_{q}(x)$ \citep{borges2004possible},
ubiquitous in one version of generalized statistical mechanics \citep{tsallis1988possible}.

The inverse function of $q-exponential$ is the $q-logarithm$ and
is given by 
\begin{equation}
ln_{q}(x)=\frac{x^{1-q}-1}{1-q},
\end{equation}
 since 
\begin{equation}
ln_{q}(e_{q}(x))=e_{q}(ln_{q}(x))=x.\label{eq:prop1}
\end{equation}

A useful relation that will be used forward can be easily proved:
\begin{equation}
[e_{q}(x)]^{a}=e_{1-\frac{(1-q)}{a}}(ax),\label{eq:prop2}
\end{equation}
 where $a$ is a constant and $q$ well known the entropic index. 

The dual conformable integral can be put in the following way \citep{rosa2018dual}:
\begin{equation}
\tilde{I}_{x}^{\alpha}f(x)=\int_{0}^{x}f(x')^{1-\alpha}f(x')dx'.
\end{equation}
An analog of fundamental theorem of standard calculus, a fundamental
theorem for $DCD$ and its related integral, can be shown as \citep{rosa2018dual}:
\begin{equation}
\tilde{D}_{x}^{\alpha}\tilde{I}_{x}^{\alpha}f(x)=f(x).\label{eq:Fundamental Theorem Dual}
\end{equation}

\subsection{q-Derivative}

The\textbf{ }$q-$derivative emerges in the context of non-additive
statistical mechanics and sets up a deformed algebra that takes into
account that the $q-$exponential is eigenfunction of $D_{(q)}$ \citep{borges2004possible,weberszpil2015connection}.
The proposed operator for $q$-derivative is given by:

\begin{eqnarray}
{\displaystyle D_{(q)}f(x)\equiv{\displaystyle \lim_{y\to x}\frac{f(x)-f(y)}{x\ominus_{q}y}}} & = & {\displaystyle [1+(1-q)x]\frac{df(x)}{dx}.}\label{eq:q-derivative}
\end{eqnarray}

Here, $\ominus_{q}$ is the deformed difference operator, $x\ominus_{q}y\equiv\frac{x-y}{1+(1-q)y}\qquad(y\ne1/(q-1)).$

There are intimate connections, see Ref. \citep{weberszpil2015connection},
between q-derivatives and conformable derivatives.

\section{Maximum Entropy Principle with Constraint of Finite $\boldsymbol{\nu}$-Moment}

In this section we prove that using a finite $\nu-moment$ and the
maximum entropy principle, an stretched exponential function emerges
as a natural and simple solution.

Following Ref. \citep{luevano2013statistical}, we impose the supplementary
condition, that the $\nu-moment$ is finite. The constraint is determinant
in this approach. 
\begin{equation}
\langle X^{\nu}\rangle=\int_{\gamma}|x^{\nu}|f(x)dx.
\end{equation}

The entropy of the system being 
\begin{equation}
S(F)=-\int_{\gamma}F(x)\ln(F(x))dx.
\end{equation}

Considering the functional 
\begin{equation}
\Phi(F)=S(F)+\alpha\left(\int_{\gamma}F(x)dx-1\right)-\beta\left(\int_{\gamma}|x^{\nu}|F(x)dx-E_{\nu}\right),
\end{equation}
and applying the variational approach, with $\delta\Phi=0$, the result
show a well known function,

\begin{equation}
F(x)=A\exp(-\beta|x^{\nu}|),
\end{equation}
that is the $KWW$ function.
\begin{equation}
A=\exp(\alpha-1).
\end{equation}
Other details can be found in the reference cited above. This is a
very more simple result when compared to that of Ref. \citep{anteneodo1999maximum}
and we indicate here the importance of the constraints to the final
result.

\section{Models of Langevin Equation with Deformed Derivatives}

Here, we pursue expressions for some general Langevin equation and
its solutions based on some the previous works with different forms
of $DD$ \citep{balankin2012hydrodynamics,balankin2012map,balankin2016towards,Jap-Weber-Sotolongo,rosa2018dual,weberszpil2015connection,weberszpil2016variational,weberszpil2017generalized,weberszpil2017structural}.

Following the proposal of Ref. \citep{sotolongo2021explicit}, the
approach with $DD$ is a simple and efficient option to obtain the
equations describing the dynamics for a broad variety of systems.
In Ref. \citep{leopoldino2019discussing}, was suggested that with
the extended variational formalism, it is possible to formulate Lagrangians
that provide the equations describing the dynamics for diverse systems,
including dissipative and non-linear systems, the description of stochastic
processes and the heat transfer equation. Also, by adopting the piece
wise form of $DD$ embedded into the Lagrangian, there was shown that
the formalism yields the correct equations of motion for such systems.

Also, we emphasize that the paradigm we adopt here \citep{Weberszpil_Helayel-Neto_2017}
is different from the standard approach in the generalized statistical
mechanics context \citep{tsallis1988possible}, where the modification
of entropy definition leads to the modification of algebra and consequently
to the concept of derivative. We analyze indirectly the influence
of a possible fractal geometry by the use of $DD$.

Within this context, it can possible justify some general results
obtained in our article. We emphasize here, that this was structured
by mapping to a continuous fractal space \citep{balankin2012map,balankin2012hydrodynamics}
which leads naturally to the necessity of modifications in the derivatives,
that gave justifications to the use of $DD$ \citep{weberszpil2015connection,weberszpil2016variational}.
The modifications of derivatives, accordingly with the metric, brings
to a change in the algebra involved, which in turn may conduct to
a generalized statistical mechanics with some adequate definition
of entropy. In short, the need for the use of the $DD$ is inter-connected
to the different degrees of freedom and also related to the reversibility/irreversibly
process; some justifying details may be found in previous publications
in Ref. \citep{weberszpil2017structural,Jap-Weber-Sotolongo,weberszpil2015connection,weberszpil2016variational,weberszpil2017generalized}.
We also claim that the use of $DD$ may enable us to consider the
effects of internal times of systems \citep{Jap-Weber-Sotolongo}.

In what concerns relating the nonlocal Fractional Calculus ($FC$)
and local operators like $DD$, it is worth mentioning similar reported
responses, e.g, with viscoelastic systems \citep{su2019fractional},
between Hausdorff derivative models and fractional derivative models,
due to the presence of a power-law into the description of viscoelastic
systems studied. The Hausdorff derivative has is in fact the same
form of conformable derivative, for differentiable function; see section
2. In what follows, we consider only the average hydrodynamic loss
between all the binary collision events. For this intention, we can
write the hydrodynamic interaction particle as deformed $LE$, but
here without considering explicitly the noise term $\eta\left(t\right)$.

In the sequence, we propose three possible dynamics description models
for velocity distribution of granular gases, using different LE modified
by different forms of $DD$.

\subsubsection{Conformable Derivative Langevin Equation}

Here by using the comformable derivative form of $DD$ for differentiable
functions, eq. (\ref{eq:conformable-deriv-differentiable}), we can
write the conformable Langevin-like equation without noise term ${\displaystyle \eta\left(t\right)}$
as
\begin{eqnarray}
D_{t}^{\alpha}v(t) & = & -\lambda v(t),\label{eq:Cinetica Conforme}
\end{eqnarray}
The solution again is obtained directly from a simple integration
and is given in terms of a stretched exponential as 
\begin{eqnarray}
v(t) & = & v_{0}e^{\frac{-\lambda t^{\alpha}}{\alpha}}.\label{eq:Sol Cinetica Conforme}
\end{eqnarray}

\subsubsection{Dual conformable Derivative Langevin Equation}

Now we follow by considering dual conformable derivative. In this
case, the Langevin-like equations without noise term ${\displaystyle \eta\left(t\right)}$assumes
the form

\begin{eqnarray}
\tilde{D}_{t}^{\alpha}v(t) & = & -\lambda v(t).\label{eq:Cinetica Dual}
\end{eqnarray}

Substituting the explicit form of $DCD$ in eq.(\ref{eq:Cinetica Dual}),
we can rewrite this, kinetic equation as 
\begin{eqnarray}
v^{\alpha-1}\frac{dv}{dt} & = & -\lambda v.
\end{eqnarray}

Note the possible equivalence with equation (26) in Ref. \citep{ben2005power},
by changing variables.

The solution of this equation can be found by simple integration and
is given in terms of $q-exponential$ as power-law velocity distribution,

\begin{eqnarray}
v(t) & = & v_{0}\left[1-(\alpha-1)\frac{\lambda t}{v_{0}^{\alpha-1}}\right]^{^{\frac{1}{\alpha-1}}},
\end{eqnarray}
that is Haff's-like law for granular gases. See and compare also eq.(1)
in Ref. \citep{yu2020velocity} and the eq.(27) in Ref. \citep{ben2005power}.

Another example of other applications of this kind of distribution
in granular systems can be found Ref. \citep{gheorghiu2003power},
where the author claim that bubbling fluidized beds are very well
fitted by a probability density function with power-law tails, following
non-Gaussian statistics, typically used in the context of the Tsallis
statistics.

Writing $v(0)=v_{0}$ and reparametrizing, by taking the entropic
parameter $q'=2-\alpha$, we obtain

\begin{eqnarray}
v(t) & = & v_{0}e_{q'=2-\alpha}\left(\frac{-\lambda t}{v_{0}^{1-q'}}\right).\label{eq:Sol Cinetica Dual}
\end{eqnarray}

Equation (\ref{eq:Sol Cinetica Dual}) indicates a different dynamics,
compatible with non-additive statistical mechanics. One interesting
point to note is that the solution depends on the initial number of
accessible states in both, multiplicative factor and the argument
of $q-exponential.$

Precedents of q-exponential appearance and similar non-Gaussian distributions
can be accessed in some article and show the consistence of the approach,
e.g., using the stochastic approach. An attempt to obtain velocity
probability distribution function $P(v)$ for a model system of granular
gas within the framework of generalized statistical mechanics, can
be found in Ref. \citep{sattin2003derivation}. Also, in Ref. \citep{combe2015experimental},
the authors explore the origins of the macroscopic friction in confined
granular materials using a q-Gaussian distribution, based on same
non-Gaussian statistics and using porous media equation. In our recent
work \citep{weberszpil2020dual}, using a variational approach with
$DCD$, we have obtained the porous medium equation and presented
insight for the solution in terms of the q-Gaussian.\vspace*{-20pt}

\subsubsection{The q-Derivative Langevin Equation}

One can write the q-Langevin-like equation without noise term ${\displaystyle \eta\left(t\right)}$
as \vspace*{-30pt}\textbf{
\begin{eqnarray}
D_{(q'')}v(t) & = & -\lambda v(t),\label{eq:q-Cinetica}
\end{eqnarray}
}\vspace*{-20pt}The solution again is obtained directly from integration
and is given in terms of a $q-exponential,$\vspace*{-20pt}
\begin{eqnarray}
v(t) & = & v_{0}[e_{q''}(t)]^{-\lambda},\label{eq:Sol-q-Cinetica}
\end{eqnarray}
with $q''$ as an entropic parameter, that in principle can be different
from $q$ or $q'.$ This solution is again very related to Tallis
non-additive entropy, since it is given in terms of the ubiquitous
$q-exponential.$

Note also that considering the mean kinetic energy, $\frac{1}{2}m<v>^{2}$,
it is possible to obtain the granular temperature distribution $<T_{g}>$
for all models using\vspace*{-20pt}

\begin{eqnarray}
<T_{g}> & = & \frac{1}{2}m<v>^{2}.
\end{eqnarray}

\begin{figure}[H]
\includegraphics[scale=0.7]{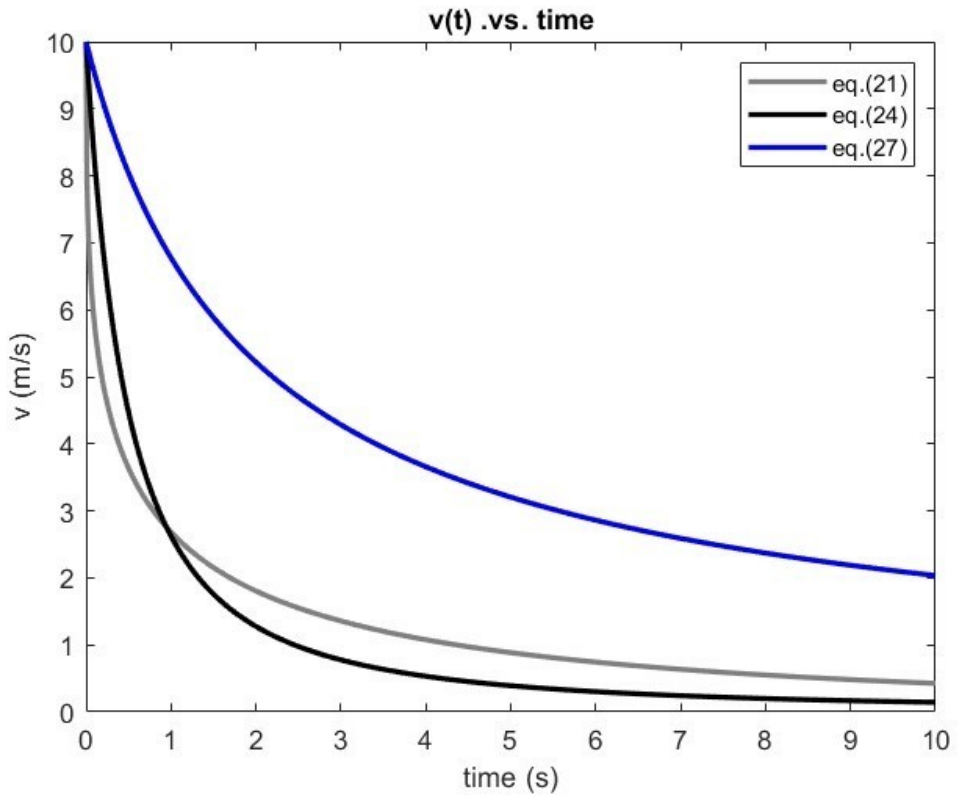}

\caption{\label{fig:LangevinSoluts}Simulations for the solutions of Langevin
equation, eqs. (21, 24 and 27). The parameters used were $v_{0}$=10m/s;
$\lambda=0.5$; $\alpha=0.38$; }
\end{figure}

See figure (\ref{fig:LangevinSoluts}) for comparisons of different
solutions.

In the sequence, we obtain a temperature dependent on specific heat
for the granular gas model using the stretched exponential distribution,
associated with the LE deformed by a $CD$.

\section{Partition Function for Stretched Exponential Probability}

In this section, considering the complexity of a granular gas, we
obtain thermodynamic quantities, based on the SEPD.

To pursue such thermodynamic results, we follow steps on Refs. \citep{plastino2017strong,plastino2018tsallis},
but here the considered distribution is the SEPD.

Using appropriate units, we can write the partition function $Z$
as 
\begin{equation}
Z=\int_{v}d^{\nu}q_{1}d^{\nu}q_{2}...d^{\nu}q_{N}d^{\nu}p_{1}d^{\nu}p_{2}...d^{\nu}p_{N}e^{-\beta{\cal {H}}^{\alpha}},
\end{equation}
where $\beta$ is the Boltzmann factor and the integration is over
generalized coordinates $q_{i}$'s and momenta $p_{i}'s$. Here, $V^{N}$
is the volume and finally, $\alpha$, the exponent of a stretched
exponential that will be better determined after some considerations.

With system's Hamiltonian $H(q_{1},...,q_{n},p_{1},...,p_{n},t)$
in a form of stretched exponential, the partition function can be
written explicitly as: 
\begin{equation}
Z=V^{N}\int_{-a}^{+a}\exp\left[-\beta\left(\dfrac{1}{2m}\right)^{\alpha}\left(p_{1}^{2}+p_{2}^{2}+...+p_{N}^{2}\right)^{\alpha}\right]dp_{1}^{\nu}dp_{2}^{\nu}...dp_{N}^{\nu}.
\end{equation}

Again, as in Ref. \citep{plastino2017strong}, using spherical coordinates
in a $\nu N$ dimensional space and calling $p_{1}^{2}+p_{2}^{2}+...+p_{N}^{2}=p^{2}$,
we obtain for $Z,$ 
\begin{equation}
Z=\dfrac{2\pi^{\nu N/2}}{\Gamma^{\nu N/2}}V^{N}\int_{0}^{\infty}\exp\left(-\dfrac{\beta}{(2m)^{\alpha}}p^{2\alpha}\right)p^{\nu N-1}dp\thinspace.
\end{equation}

In order to solve the integral, we can make a change of variable: 

\begin{equation}
u=\dfrac{\beta}{(2m)^{\alpha}}p^{2\alpha}\longrightarrow p=\left[\dfrac{(2m)^{\alpha}}{\beta}u\right]^{1/{2\alpha}}
\end{equation}
 and 
\begin{equation}
du=\dfrac{\beta}{(2m)^{\alpha}}2\alpha p^{2\alpha-1}dp\thinspace.
\end{equation}

By this way, we can write $dp$ as 
\begin{equation}
dp=\dfrac{(2m)^{\alpha}}{2\alpha\beta}p^{1-2\alpha}du=\dfrac{(2m)^{\alpha}}{2\alpha\beta}\left[\dfrac{(2m)^{\alpha}}{\beta}u\right]^{\dfrac{{1-2\alpha}}{2\alpha}}du.
\end{equation}

The partition function, $Z$, can now be written as

\begin{eqnarray}
Z & = & \dfrac{\pi^{\nu N/2}}{\Gamma^{\nu N/2}}V^{N}\int_{0}^{\infty}e^{-u}\left[\dfrac{(2m)^{\alpha}}{\beta}u\right]^{(2\nu-1)/{2\alpha}}\dfrac{(2m)^{\alpha}}{2\alpha\beta}\left[\dfrac{(2m)^{\alpha}}{\beta}u\right]^{\dfrac{{1-2\alpha}}{2\alpha}}du=\nonumber \\
 & = & \dfrac{\pi^{\nu N/2}}{\Gamma^{\nu N/2}}V^{N}\left[\dfrac{(2m)^{\alpha}}{\beta}\right]^{(2\nu-1)/{2\alpha}}\dfrac{(2m)^{\alpha}}{2\alpha\beta}\left[\dfrac{(2m)^{\alpha}}{\beta}\right]^{\dfrac{{1-2\alpha}}{2\alpha}}\int_{0}^{\infty}e^{-u}u^{(2\nu-1)/{2\alpha}}u^{1-2\alpha/2\alpha}du=\nonumber \\
 & = & C_{1(\nu,N,\alpha,m,\beta)}\int_{0}^{\infty}e^{-u}u^{(\nu N-1+1-2\alpha)/{2\alpha}}du=C_{1(\nu,N,\alpha,m,\beta)}\int_{0}^{\infty}e^{-u}u^{(\nu N-2\alpha)/{2\alpha}}du.\label{eq:Z1}
\end{eqnarray}

In the eq.(\ref{eq:Z1}) we can identified the form of a gamma function
$\Gamma$ as
\begin{equation}
\Gamma(u)=\int_{0}^{\infty}e^{-u}u^{(\frac{\nu N}{2\alpha}-1)}du.
\end{equation}

Following with the calculations, the pre factor in eq.(\ref{eq:Z1})
can be written, with some simple algebraic as 
\begin{eqnarray}
C_{1} & \equiv & C_{1(\nu,N,\alpha,m,\beta)}=\dfrac{(2m)^{{{(\nu N-2\alpha})/{2}}}}{\beta^{{(\nu N-2\alpha})/{2\alpha}}}\dfrac{(2m)^{\alpha}}{2\beta\alpha}\left(\dfrac{\pi^{\nu N/2}}{\Gamma(\nu N/2)}V^{N}\right)=\nonumber \\
 & = & \dfrac{(2m)^{{{(\nu N})/{2}}}}{2\alpha\beta^{{(\nu N})/{2\alpha}}}\left(\dfrac{\pi^{\nu N/2}}{\Gamma(\nu N/2)}V^{N}\right).
\end{eqnarray}

Finally, $Z$ can be cast as\vspace*{-20pt}

\begin{equation}
Z=\dfrac{(2m)^{{{(\nu N})/{2}}}}{2\alpha\beta^{{(\nu N})/{2\alpha}}}\left(\pi^{\nu N/2}V^{N}\right).\label{eq:Partition Func 1}
\end{equation}

Note that 

\begin{equation}
\dfrac{\nu N}{2\alpha}>0\Longrightarrow\nu N>0\Longrightarrow N>0,
\end{equation}
for acceptable physical solution and without the restrictions of poles
in Plastino\textquoteright s papers \citep{plastino2017strong,plastino2018tsallis}.

To calculate the mean energy $\langle U\rangle,$we proceed in a similar
way, following Refs. \citep{plastino2017strong,plastino2018tsallis}:

\begin{equation}
Z\ast\langle U\rangle=V^{N}\int_{-\infty}^{+\infty}\exp\left[-\dfrac{\beta}{(2m)^{\alpha}}\left(p_{1}^{2}+p_{2}^{2}+...+p_{N}^{2}\right)^{\alpha}\right]\dfrac{\left(p_{1}^{2}+p_{2}^{2}+...+p_{N}^{2}\right)}{2m}d^{\nu}p_{1}d^{\nu}p_{2}...d^{\nu}p_{N}.
\end{equation}

Integrating over the angles we find that,

\begin{equation}
Z\ast\langle U\rangle=\dfrac{2\pi^{\nu N/2}}{(2m)\Gamma\left(\dfrac{\nu N}{2}\right)}\int_{0}^{\infty}\exp\left(\dfrac{-\beta}{(2m)^{\alpha}}p^{2\alpha}\right)p^{\nu N+1}dp.
\end{equation}

Changing conveniently the variable $p,$ we can write

\begin{equation}
u=\dfrac{\beta}{(2m)^{\alpha}}p^{2\alpha}\Longrightarrow p=\left[\dfrac{(2m)^{\alpha}u}{\beta}\right]^{1/{2\alpha}},
\end{equation}

and

\begin{equation}
du=\dfrac{\beta}{(2m)^{\alpha}}2\alpha p^{2\alpha-1}dp.
\end{equation}

So, $dp$ and $p^{\nu N+1}$ can be written as

\begin{eqnarray}
dp & = & \dfrac{(2m)^{\alpha}}{2\alpha\beta}p^{1-2\alpha}du=dp=\dfrac{(2m)^{\alpha}}{2\alpha\beta}\left[\dfrac{(2m)^{\alpha}u}{\beta}\right]^{{(1-2\alpha)}/{2\alpha}}du,\nonumber \\
p^{\nu N+1} & = & \left[\dfrac{(2m)^{\alpha}u}{\beta}\right]^{(\nu N+1)/{2\alpha}}.
\end{eqnarray}

The product $Z\ast\langle U\rangle$ can be written as

\begin{equation}
Z\ast\langle U\rangle=\dfrac{V^{N}\pi^{\nu N/2}}{(2m)\Gamma\left(\dfrac{\nu N}{2}\right)}\int_{0}^{\infty}e^{-u}\left[\dfrac{(2m)^{\alpha}}{\beta}u\right]^{(\nu N+1)/{2\alpha}}\dfrac{(2m)^{\alpha}}{2\alpha\beta}\left[\dfrac{(2m)^{\alpha}u}{\beta}\right]^{(1-2\alpha)/{2\alpha}}du,
\end{equation}

\begin{equation}
Z\ast\langle U\rangle=C_{2(\nu,N,\alpha,m,\beta)}\int_{0}^{\infty}e^{-u}u^{(\nu N+1)/{2\alpha}}u^{(1-2\alpha)/{2\alpha}}du,
\end{equation}
where $C_{2}$ is the pre factor with terms not containing $u.$

\begin{eqnarray}
Z\ast\langle U\rangle & = & C_{2(\nu,N,\alpha,m,\beta)}\int_{0}^{\infty}e^{-u}u^{(\nu N+2-2\alpha)/{2\alpha}}du=\nonumber \\
 & = & C_{2}\int_{0}^{\infty}e^{-u}u^{(\nu N+2)/{2\alpha}-1}du.
\end{eqnarray}

Explaining the mean energy, we have

\begin{equation}
\langle U\rangle=\frac{C_{2(\nu,N,\alpha,m,\beta)}}{Z}\Gamma\left(\dfrac{\nu N+2}{2\alpha}\right).\label{eq:Mean Energy1}
\end{equation}

Note that the argument of the gamma function is positive, that is
\begin{equation}
\dfrac{\nu N+2}{2\alpha}>0.
\end{equation}
In this way, it follows that $\nu N+2>0$ $\Longrightarrow$ $N>-2/{\nu}$
$\Longrightarrow$ $N>0$, for physical solutions. Again, there are
no the restrictions of poles (see Refs. \citep{plastino2017strong,plastino2018tsallis}). 

After some simple algebraic, the term $C_{2}$ can be written as

\begin{equation}
C_{2(\nu,N,\alpha,m,\beta)}=\dfrac{V^{N}}{2\alpha}\dfrac{(2m)^{(\nu N)/{2}}}{\beta^{(\nu N+2)/{2\alpha}}}\dfrac{\pi^{\nu N/2}}{\Gamma\left(\dfrac{\nu N}{2}\right)}.
\end{equation}

Using a correction factor \citep{magomedov2018generalization} 
\begin{equation}
\frac{1}{N!(2\pi\hbar)^{\nu N}},
\end{equation}
we can rewrite the term $C_{2}$ as: 

\begin{equation}
C_{2}\equiv C_{2(\nu,N,\alpha,m,\beta)}=\dfrac{1}{N!(2\pi\hbar)^{\nu N}}\dfrac{V^{N}}{2\alpha}\dfrac{(2m)^{(\nu N)/{2}}}{\beta^{(\nu N+2)/{2\alpha}}}\dfrac{\pi^{\nu N/2}}{\Gamma\left(\dfrac{\nu N}{2}\right)}.
\end{equation}

Finally, the Helmholtz free energy can also be written as:

\begin{equation}
F=-kT\ln Z_{\alpha}=-KT\left[-lnN!-\ln(2\pi\hbar)^{\nu N}+\ln\left(\pi^{\nu N/2}\right)+\ln\left(\dfrac{2m^{nuN})}{2\beta^{\nu/2}\alpha}\right)+NlnV\right].
\end{equation}

\subsubsection{The State Equation}

Following standard thermodynamic relation, an unchanged state equation
in obtained,

\begin{equation}
P=\dfrac{\partial F}{\partial V}=-KT\dfrac{N}{V}.
\end{equation}

The expression above can indicate that the relation between Helmholtz
free energy $F$ and the volume $V$could be related by another king
of derivative, like deformed derivative \citep{weberszpil2017generalized}.

Despite the unchanged state equation, the mean energy or internal
energy expression, in eq. (\ref{eq:Mean Energy1}) changed, as was
shown above.

\subsection{The Internal Energy and Temperature-Dependent Specific Heat }

Let us write explicitly the mean internal energy. With eq. (\ref{eq:Mean Energy1})
and eq.(\ref{eq:Partition Func 1}) written as $Z=C_{1}\Gamma\left(\dfrac{\nu N}{2\alpha}\right),$the
internal mean energy can be written as

\begin{equation}
\langle U\rangle=\dfrac{C_{2}}{C_{1}}\dfrac{\Gamma\left(\dfrac{\nu N+2}{2\alpha}\right)}{\Gamma\left(\dfrac{\nu N}{2\alpha}\right)},
\end{equation}
with $C_{1}=\dfrac{(2m)^{{{(\nu N})/{2}}}}{2\alpha\beta^{{(\nu N})/{2\alpha}}}\left(\dfrac{\pi^{\nu N/2}}{\Gamma(\nu N/2)}V^{N}\right).$
By this way, $\dfrac{C_{2}}{C_{1}}$ is identified as

\begin{equation}
\dfrac{C_{2}}{C_{1}}=\dfrac{\beta^{\nu N/2}}{\beta^{(\nu N+2)/2}}=\beta^{\dfrac{\nu N-\nu N-2}{2\alpha}}=\beta^{-1/\alpha}.
\end{equation}

The internal mean energy of the granular gas is then explained as

\begin{equation}
\langle U\rangle=\beta^{-1/\alpha}\dfrac{\Gamma\left(\dfrac{\nu N+2}{2\alpha}\right)}{\Gamma\left(\dfrac{\nu N}{2\alpha}\right)}\thinspace.\label{Internal Energy}
\end{equation}

It is possible to give a one parametric equation for the specific
heat $C_{V}$, that we will show in what follows that depends on the
temperature.

Performing a partial derivative on the mean energy, the $C_{V}$ can
now be written as

\begin{equation}
C_{V}=\dfrac{\partial\langle U\rangle}{\partial T}=\dfrac{\partial}{\partial T}\left(\beta^{-1/\alpha}\dfrac{\Gamma\left(\dfrac{\nu N+2}{2\alpha}\right)}{\Gamma\left(\dfrac{\nu N}{2\alpha}\right)}\right),
\end{equation}
or

\begin{equation}
C_{V}=\dfrac{K^{1/\alpha}}{\alpha}T^{(1-\alpha)/\alpha}\dfrac{\Gamma\left(\dfrac{\nu N+2}{2\alpha}\right)}{\Gamma\left(\dfrac{\nu N}{2\alpha}\right)},\label{eq:CV}
\end{equation}
where $K$ is the Boltzmann constant. The parameter $\alpha$ is an
experimental fractal parameter and depend on the granular gas composition.

To perform the simulation, we will use the Stirling's formula, that
allows derivation of the following asymptotic expansion for the ratio
of gamma functions:

\begin{equation}
\frac{\Gamma(x+c)}{\Gamma(x)}\approx x^{c}.
\end{equation}

With Stirling approximation, the equation (\ref{eq:CV}) can be rewritten
as 
\begin{equation}
C_{V}\approx\dfrac{K^{1/\alpha}}{\alpha}T^{(1-\alpha)/\alpha}\left(\dfrac{\nu N}{2\alpha}\right)^{1/\alpha}
\end{equation}

In order to better emulate the nature of a real gas, let us redefine
the parameter $\alpha$ as $\alpha=2-\eta.$ The equation (\ref{eq:CV})
can be written now as

\begin{equation}
C_{V}\approx\dfrac{K^{1/(2-\eta)}}{2-\eta}T^{(\eta-1)/(2-\eta)}\left(\dfrac{\nu N}{2-\eta}\right)^{\frac{1}{(2-\eta)}}.\label{eq:CvStirling-Eta}
\end{equation}

The reader can observe that in the Ref. \citep{magomedov2018generalization}
(Table 1, column 3), the specific heat at constant volume, $C_{V}$,
for the noble gas Argon ($Ar$), present a low fractionality, $\alpha$,
that is, $\alpha$ is very close to one. This reinforce that the adequate
approach for the thermodynamics of granular systems, like noble gases,
is not based on non local fractional calculus ($FC$) but instead,
can be better described by fractal (or named deformed) derivatives,
that are local an simpler than $FC$ operators.

Here, $\alpha,\eta$ are dimensionless parameter only. 

In Ref. \citep{mamedov2018evaluation} some values of $C_{V}$ are
shown.

\begin{figure}[t]
\includegraphics[scale=0.6]{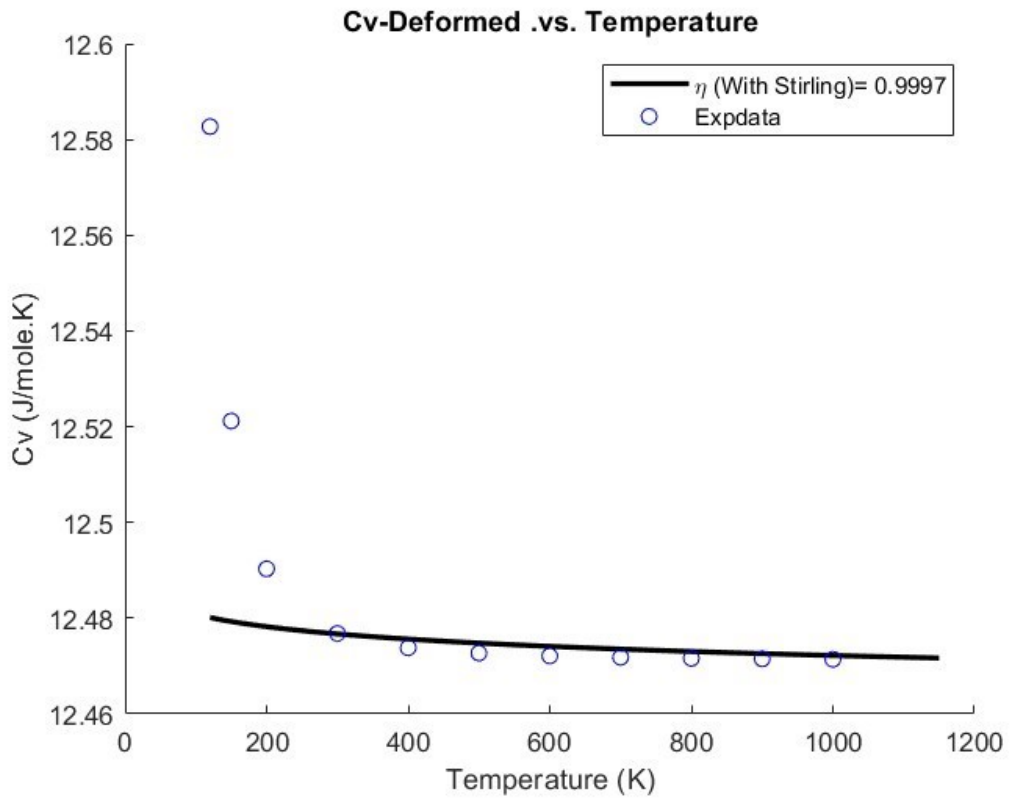}

\caption{\label{fig:.Cv-VersusTemp}$C_{V}$.vs. $T$ for different values
of parameter $\eta.$ Experimental data for Argon (Ar) by Ref. \citep{mamedov2018evaluation}.}
\end{figure}

In Figure (\ref{fig:.Cv-VersusTemp}) we plotted the experimental
data obtained from Ref. \citep{mamedov2018evaluation} and adjusted
via eq. (\ref{eq:CvStirling-Eta}) for $C_{V}.$As it can be seen,
the fitting shows a good agreement with experimental data for temperatures
above $300K$. For low temperatures, a quantum model has to be adapted. 

\begin{figure}[H]
\includegraphics[scale=0.6]{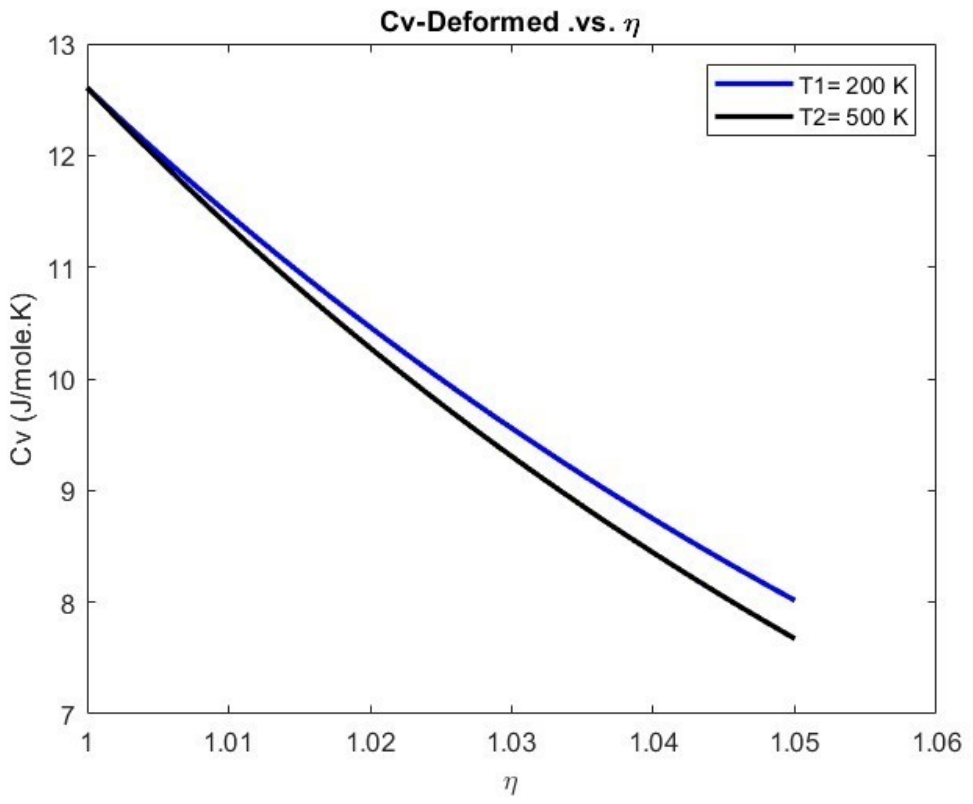}

\caption{\label{fig:.Cv-Versus-Eta}$C_{V}$.vs. $\eta$ at different temperatures
(K).}
\end{figure}

In Figure \ref{fig:.Cv-Versus-Eta} the reader can see a plotting
of $C_{V}$ versus parameter $\eta$, for temperatures of $200K$
and $500K$.

\subsection{Consistence with classical approach of kinetic theory}

Let us now examine the consistence of the relations here obtained
with those of classical kinetic theory approach. For this intention
let us make the parameter $\alpha$ tends to $1,$ $\alpha\rightarrow1:$
In this case

\begin{eqnarray}
C_{v} & \rightarrow & K\dfrac{\Gamma\left(\dfrac{\nu N+2}{2}\right)}{\Gamma\left(\dfrac{\nu N}{2}\right)}=\nonumber \\
 & = & K\dfrac{\Gamma\left(\dfrac{\nu N}{2}+1\right)}{\Gamma\left(\dfrac{\nu N}{2}\right)}=\nonumber \\
 & = & K\dfrac{\nu N}{2}\dfrac{\Gamma\left(\dfrac{\nu N}{2}\right)}{\Gamma\left(\dfrac{\nu N}{2}\right)}=\nonumber \\
 & = & K\dfrac{\nu N}{2},
\end{eqnarray}
where we have used that, for gamma function, is valid the relation
$\Gamma\left(x+1\right)=x\times\Gamma\left(x\right).$

For $3D$ gas, that is, for $\nu=3$, the specific heat turns out
to be the well known classical one for the ideal gas,

\begin{equation}
C_{v}=\dfrac{3}{2}NK,
\end{equation}
and the internal mean energy is

\begin{equation}
\langle U\rangle=KT\dfrac{\nu N}{2}.
\end{equation}
 When $\nu=3,$

\begin{equation}
\langle U\rangle=\dfrac{3}{2}NKT,
\end{equation}
that is the well known equipartition principle for classical ideal
gases. 

\section{Conclusions and Outlook For Further Investigations}

In this contribution we begin by confirming that when we use constraints
with finite momentum and the principle of maximum entropy, the Kohlrausch--Williams--Watts
function, known as stretched exponential, emerges naturally and in
a simpler way, when compared to results in the literature and following
according to Ref.\citep{luevano2013statistical}. 

Using the $DD$ mathematical tool, we obtain generalized expressions
for the Langevin equation, as well as its solutions for three different
forms of $DD$, including those connected with the so useful nonaddictive
statistical mechanics. The Haff's-like law for granular gases and
its solutions were obtained. We call attention for the Ref. \citep{Cresus2024Remarks}
with a generalized thermodynamics and quantum models.

In terms of the stretched exponential function, we calculate the partition
function $Z$ for a granular gas system by building up the probability
density and, from this partition function, we determine the internal
energy of the system as well as the specific heat $C_{V}$, both dependent
on temperature $T$. Comparisons with experimental data from \citep{mamedov2018evaluation}
for Argon have shown good accordance above $250K.$ For low temperature
experiments, a quantum model is necessary as in Einstein-like model. 

In Ref. \citep{Cresus2024Remarks} we are developing some quantum
models for a generalized thermodynamics and will be published elsewhere.

The consistency with classical approach of kinetic theory for ideal
gases was verified. 

We emphasize that the adequate approach for thermodynamics (not quantum
thermodynamics) of granular systems, like noble gases, is not based
on non local $FC$ but instead, can be better described by deformed
derivatives in its diverse forms, that are local an simpler than $FC$
operators, due to the low fractionality of these systems \citep{magomedov2018generalization}.

\textbf{Acknowledgments: }The authors wish to express their gratitude
to FAPERJ, APQ1, for the partial financial support.

\bibliographystyle{aapmrev4-2}
\bibliography{reference}
 
\end{document}